# 基于多源域适应的缺陷类别预测方法

邢颖[1,2], 赵梦赐[1,2], 杨斌[3], 张俞炜[4], 李文瑾[5], 顾佳伟[5], 袁军[5]

[1](北京邮电大学 人工智能学院,北京 100876)

[2](高安全系统的软件开发与验证技术工业和信息化部重点实验室(南京航空航天大学),南京 211106)

[3](中国联通研究院,北京 100048)

[4](中国科学院 软件研究所,北京 100190)

[5](绿盟科技集团股份有限公司,北京 100089)

通讯作者: 张俞炜, E-mail: zhangyuwei@otcaix.iscas.ac.cn

**摘 要**: 随着规模和复杂性的迅猛膨胀,软件系统中不可避免地存在缺陷.近年来,基于深度学习的缺陷预测技术成为软件工程领域的研究热点.该类技术可以在不运行代码的情况下发现其中潜藏的缺陷,因而在工业界和学术界受到了广泛的关注.然而,已有方法大多关注方法级的源代码中是否存在缺陷,无法精确识别具体的缺陷类别,从而降低了开发人员进行缺陷定位及修复工作的效率.此外,在实际软件开发实践中,新的项目通常缺乏足够的缺陷数据来训练高精度的深度学习模型,而利用已有项目的历史数据训练好的模型往往在新项目上无法达到良好的泛化性能.因此,本文首先将传统的二分类缺陷预测任务表述为多标签分类问题,即使用 CWE(common weakness enumeration)中描述的缺陷类别作为细粒度的模型预测标签.为了提高跨项目场景下的模型性能,本文提出一种融合对抗训练和注意力机制的多源域适应框架.具体而言,该框架通过对抗训练来减少域(即软件项目)差异,并进一步利用域不变特征来获得每个源域和目标域之间的特征相关性.同时,该框架还利用加权最大均值差异作为注意力机制以最小化源域和目标域特征之间的表示距离,从而使模型可以学习到更多的域无关特征.在本文构建的包含 8 个真实世界开源项目的数据集上的实验表明,所提方法对比最先进的基线方法取得了显著的性能提升.

## Defect Category Prediction Based on Multi-Source Domain Adaptation

XING Ying[1,2], ZHAO Meng-Ci[1,2], YANG Bin[3], ZHANG Yu-Wei[4], LI Wen-Jin[5], GU Jia-Wei[5], YUAN Jun[5]

[1](School of Artificial Intelligence, Beijing University of Posts and Telecommunications, Beijing 100876, China)

[2](Key Laboratory for Safety-Critical Software Development and Verification (Nanjing University of Aeronautics and Astronautics), Ministry of Industry and Information Technology, Nanjing 211106, China)

[3](China Unicom Research Institute, Beijing 100048, China)

[4](Institute of Software, Chinese Academy of Sciences, Beijing 100190, China)

[5](NSFOCUS Technologies Group Co., Ltd., Beijing 100089, China)

**Abstract**: With the rapid expansion of scale and complexity, defects inevitably exist within software systems. In recent years, defect prediction techniques based on deep learning have become a prominent research topic in the field of software engineering. These techniques can identify potential defects without executing the code, garnering significant attention from both industry and academia. However, existing approaches mostly concentrate on determining the presence of defects at the method-level code, lacking the ability to precisely classify specific defect categories. Consequently, this undermines the efficiency of developers in locating and rectifying defects. Furthermore, in practical software development, new projects often lack sufficient defect data to train high-accuracy deep learning models. Models trained on historical data from existing projects frequently struggle to achieve satisfactory generalization performance on new projects. Hence, this paper initially reformulates the traditional binary defect prediction task into a multi-label classification problem, employing defect categories described in the Common Weakness Enumeration (CWE) as fine-grained predictive labels. To enhance the model performance in cross-project scenarios, this paper proposes a multi-source domain adaptation framework that integrates adversarial training and attention mechanisms. Specifically, the proposed framework employs adversarial training to mitigate domain (i.e., software projects) discrepancies, and further utilizes domain-invariant features to capture feature correlations between each source domain and the target domain. Simultaneously, the proposed framework employs a weighted maximum mean discrepancy as an attention mechanism to minimize the representation distance between source and target domain features, facilitating model in learning more domain-independent features. The experiments on the dataset consisting of 8 real-world open-source projects constructed in this paper show that the proposed approach achieves significant performance improvements compared to state-of-the-art baselines.

随着软件规模和复杂性的急剧增加,软件缺陷逐渐成为导致计算机系统失效和停机的首要原因.这些缺陷可能造成严重的财务损失,甚至危及生命安全.因此,如何在软件开发的早期阶段挖掘出软件代码中潜在的缺陷是一个亟需解决的问题.软件缺陷预测[1-3]是保证软件质量的有效手段之一,基于机器学习或深度学习的智能化技术能够在较短的时间内发现更多的缺陷模块,减少开发人员维护软件的成本.早期的研究人员利用数据驱动的方法从软件源代码中提取与软件缺陷强相关的静态度量特征,通过构建缺陷数据集来训练机器学习模型以预测目标程序代码中的缺陷数量.随着深度学习技术的发展,研究人员开始利用神经网络从复杂的代码结构中自动挖掘深层的语法语义表征,从而进一步提高缺陷预测模型的性能.

然而,已有方法通常关注给定的函数或文件代码中是否存在缺陷(即将缺陷预测问题视为一个二分类任务),但无法分析更详细的软件缺陷信息(如类别).由缺陷管理实践可知,一个软件缺陷从发现到被处理完成需要经历特定的流程,而成功修复被报告的缺陷在实际软件开发过程中是一个复杂的任务[4].因此,预测具体的缺陷类别可以为开发人员分析缺陷特征提供有益的信息,从而提高开发人员进行缺陷定位及修复工作的效率[5].为了在实际软件开发过程中减少查找、修复缺陷的成本,本文将传统的二分类缺陷预测任务表述为面向缺陷类别的多标签分类问题,使用 CWE(common weakness enumeration)[6]作为细粒度的模型预测标签.CWE 列表中包含了近千种具有安全影响的软件和硬件缺陷类别,它通过开源社区开发的模式来维护并确保列表中的每个缺陷类别都得到了充分的描述和区分.通过为检测到的软件缺陷自动预测细粒度的 CWE 缺陷类别,开发人员可以将其作为索引在 CWE 网站上检索相关的缺陷信息(如缺陷的严重程度).这些信息能够针对性地帮助开发人员采取相应的策略来修复缺陷.Zou 等人[7]首次在漏洞检测领域进行细粒度的漏洞类型分类研究,该方法通过利用深度学习模型来捕获代码语义,从而有效地进行多类漏洞检测.然而,该方法使用的训练数据大多是使用已知的缺陷模式来人工合成的,可能与真实世界缺陷数据分布有所差异,且无法捕获真实世界缺陷的复杂性[8].

在实际软件开发实践中,开发人员往往需要对新的项目进行缺陷类别预测.这些项目通常缺乏足够的缺陷标注数据,难以通过有监督的方式来训练出高效预测缺陷类别的分类模型.因此,研究人员提出了跨项目缺陷预测(cross-project defect prediction, CPDP)[9,10],即利用其他有标签项目(源项目)的历史缺陷数据来训练模型,并在新的无标签项目(目标项目)上进行预测.然而,由于不同项目在软件功能、编程语言和开发人员方面存在差异,因此构建的数据集之间往往存在数据分布上的差异,从而导致 CPDP 模型的实际性能并不理想[11].此外,已有方法[5,12,13]仅仅使用单一的源项目或直接将所有源项目视为一个源域来构造数据集训练模型以学习领域知识.这样的数据预处理方式可能会面临模型过拟合或泛化能力差的问题.因此,如何有效迁移源项目的知识来提高跨项目缺陷类别预测模型的性能也是一个有挑战性的问题.

针对上述问题,本文提出一种融合对抗训练和注意力机制的多源域适应框架 COPILOT(defe**C**t categ**O**ry **P**red**I**ction based on mu**L**ti-source d**O**main adap**T**ation)来训练缺陷类别预测模型.由于来自不同应用领域的开源软件项目具备知识多样性,本文将不同的软件项目视为不同的域.为了有效解决多个源域(源项目)向一个目标域(目标项目)迁移知识的问题,本文引入多源域适应(multi-source domain adaptation, MDA)[14]的思想.MDA 是迁移学习领域的一个重要分支,已经成功应用于模式识别、自然语言处理等领域[15,16].如图 1 所示,其核心思想是学习不同源域和目标域之间的共性,或适配它们之间的概率分布,从而使得在多源数据集上训练的模型能在目标项目数据集上具有良好的泛化性能.基于 MDA 思想,本文所提出的框架 COPILOT 旨在通过最小化不同源项目和目标项目数据集之间的数据分布差异来充分利用源项目中已标记的历史缺陷数据,从而缓解目标数据集中标注数据不足的问题,并且能够实现不同软件项目之间知识的高效迁移以提升跨项目场景下缺陷类别预测模型的分类性能.具体来说,COPILOT 首先通过对抗训练的方式提取源域和目标域的共有信息特征,并生成不同源域和目标域之间的域相关性,以确保每个源域和目标域之间的数据分布一致.接着,COPILOT 选择加权最大均值差异(weighted maximum mean discrepancy, WMMD)[17]作为注意力机制来进一步地最小化不同源域和目标域特征之间的表示距离.WMMD 利用对抗训练中获得的域相关性来对不同的源域进行加权处理,并通过权重来影响预测模型的参数学习过程,从而将不同源域中的域私有特征更好地适配到目标域中.

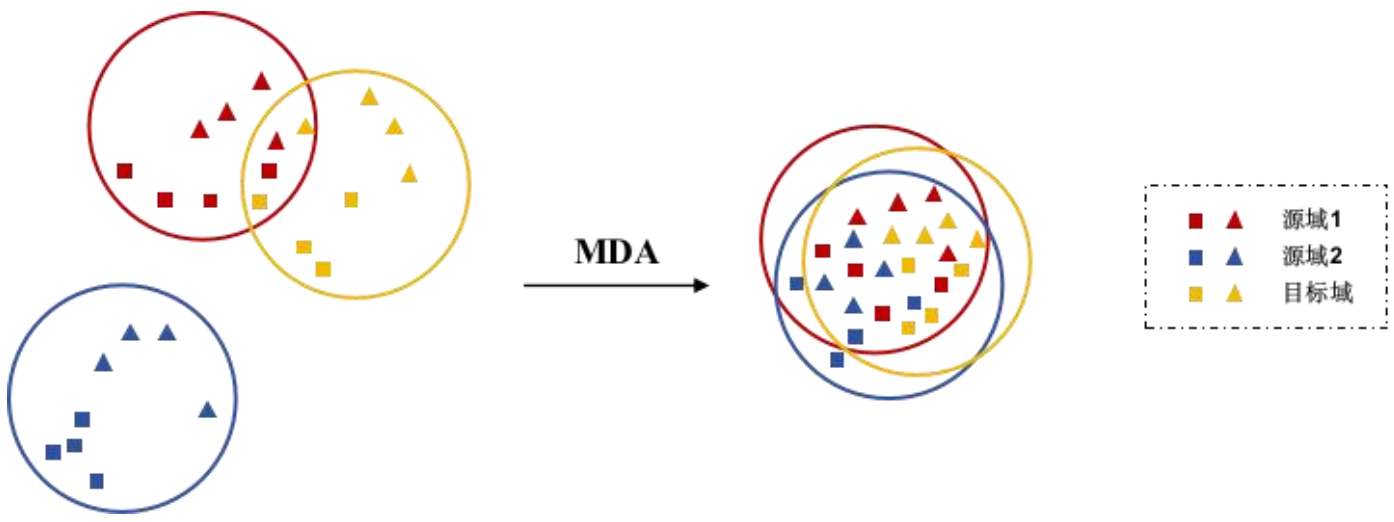

图 1　MDA 原理图(以两个源域为例)

本文第 1 节对相关工作进行介绍.第 2 节对本文提出的缺陷类别预测方法 COPILOT 进行详细介绍.第 3 节对本文采取的实验设置进行描述,并提出三个相关的研究问题.第 4 节针对三个研究问题,对实验结果与分析进行介绍.第 5 节回顾了本文的内容并对未来可能的研究方向进行展望.

## 1　相关工作

本节将对现有的缺陷类别预测方法和常见的多源域适应技术进行回顾与分析.

### 1.1　缺陷类别预测

软件缺陷类别预测是一个重要且具有挑战性的任务.在理想情况下,开发人员通常希望缺陷检测工具不仅要能判断软件代码中是否包含缺陷,还要能精确定位相关缺陷的类别.早期的缺陷预测研究[1-3]集中于利用机器学习或深度学习技术从历史软件项目仓库中自动挖掘软件代码和潜在缺陷之间的关系,从而检测缺陷的存在,但无法精确分析具体的缺陷类别.与传统的缺陷预测方法相比,面向细粒度标签的缺陷类别预测任务成为近年来的研究热点,目的是通过提供与缺陷相关的具体信息来提高开发人员后续进行软件维护工作的效率.Zou 等人[7]首次提出了基于深度学习的多类别漏洞检测方法μVulDeePecker,该方法的核心思想是代码关注度的概念,可以有助于捕捉精确定位漏洞类别的信息.Xing 等人[12]针对跨项目缺陷预测任务提出了一种采用对抗学习思想的迁移方法 AC-GAN(abstract continuous generative adversarial network).该方法通过使用生成式对抗网络来改变目标项目特征的分布,使其接近于源项目特征的分布,从而提升跨项目缺陷预测的性能.与上述方法相比,本文通过引入域适应的思想来有效利用不同源项目中已标记的历史缺陷数据,并且能够最小化不同源项目与目标项目之间的数据分布差异以缓解模型负迁移问题.

### 1.2　多源域适应

域适应[18]是迁移学习领域的一个持续研究热点.该问题的基本假设是源域和目标域的类别空间以及特征空间一致,但数据分布不一致,通过利用带标签的源域数据来学习目标域数据的标签.针对跨项目缺陷预测问题,Chen 等人[13]提出一种有监督的单源域适应方法.该方法将领域适配与机器学习模型训练过程相融合,通过构建与目标域数据集相关的权重,并将这些权重应用于源域数据集中来影响模型的参数学习过程,从而将来自目标域数据集的分布特性适配到源域数据集中.但当源域的训练数据量不足且不平衡时,单源域适应的方法容易产生过拟合的问题.为了在有多个源域的场景下进行有效的知识迁移,本文引入多源域适应技术来减小不同源域与目标域之间的表示距离.为了缓解单源域适应算法在多源域适应场景下可能会导致次优解的问题,Zhao 等人[19]提出了一种多源域对抗网络模型 MDAN(multi-source domain adversarial network),通过对抗训练来优化域适应的泛化约束,从而产生更具数据效率和任务适应性的模型.Peng 等人[20]收集并标注了迄今为止最大的用于多源域适应任务的数据集 DomainNet,并提出了新的基于矩匹配的多源域适应方法 M3SDA(moment matching for multi-source domain adaptation),旨在通过动态对齐特征分布的矩,将从多个已标注的源域学到的知识迁移到未标注的目标域.Zuo 等人[21]提出了一种基于注意力的多源域适应方法 ABMSDA(attention-based multi-source domain adaptation),通过考虑域相关性来减轻异域带来的影响.为了获得源域和目标域之间的域相关性,ABMSDA 首先训练域识别模型来计算目标域数据样本属于每个源域的概率.根据域相关性,该方法进一

步提出加权矩距(weighted moment distance, WMD)以更多地关注相似性较高的源域.与上述方法相比,本文提出一种融合对抗训练和注意力机制的多源域适应框架 COPILOT.该方法首先通过对抗训练的方式学习不同源域和目标域之间的域相关性,然后使用基于域相关性的加权最大均值差异作为注意力机制来将不同源域中的域私有特征更好地适配到目标域中,从而达到最小化不同源域与目标域之间的数据分布差异的目的.

## 2　基于多源域适应的缺陷类别预测方法 COPILOT

为了进行缺陷类别预测,本文提出了一种端到端的多源域适应框架 COPILOT.如图 2 所示(为了便于理解,图中以两个源域和一个目标域为例进行展示),该框架主要包括三个模块: **1)域对抗训练:** 首先利用域编码器对不同源域和目标域中的样本进行特征提取,然后将编码后的特征表示经过梯度反转后输入域鉴别器以生成不同源域与目标域之间的相关性权重矩阵; **2)加权最大均值差异:** 首先利用特征编码器对不同源域和目标域中的样本进行特征提取,然后将不同源域与目标域的特征表示分别计算最大均值差异,最后将对抗训练得到的域相关性权重矩阵对特征表示进行进一步加权; **3)缺陷类别预测:** 利用经过多源域适应处理的特征表示训练分类器以预测缺陷类别.本节将首先对 COPILOT 框架中的组件架构进行简要介绍,并针对缺陷类别预测问题给出相关符号定义.

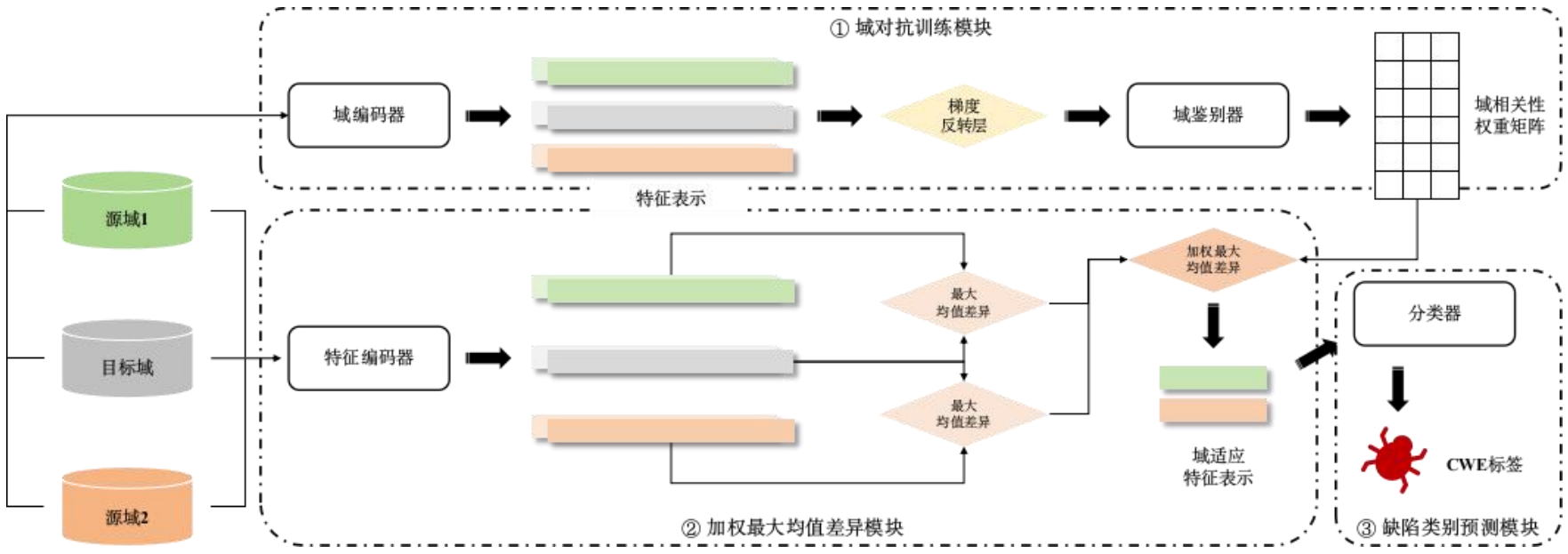

图 2　COPILOT 框架概述

### 2.1　组件架构及符号定义

如图 2 所示,本文使用基于大规模源代码进行预训练的语言模型 CodeT5[22]作为 COPILOT 框架中的编码器组件(即域编码器与特征编码器).CodeT5 是基于 Transformer 架构[23]的、广泛用于源代码理解领域的编码器-解码器模型,可以很好地支持源代码相关的推理与生成任务[24,25].此外,框架中的域鉴别器以及分类器两个组件则是使用全连接层作为分类层,并利用 softmax 函数来分别计算每个分类标签的概率.针对缺陷类别预测任务,本文将采样自不同软件项目中的缺陷数据样本认为是来自不同的领域.在多源域适应场景下,不同源域表示已有的已标记缺陷样本充足的源项目数据集,目标域则表示缺陷样本标签需要预测的目标项目数据集.如图 2 所示,给定两个源域( $D^{S_1}$ 和 $D^{S_2}$ )以及目标域 $D^{\mathrm{T}}$ , $D^{S_i}$ 和 $D^{\mathrm{T}}$ 具有相同的缺陷类别标签,但样本和类别标签的联合概率分布不同.令 $D^{S_i}=\left\{\left(X_j^{S_i},Y_j^{S_i}\right)\right\}_{j=1}^{n_{S_i}}$ ,其中 $n_{S_i}$ 表示第 i 个源域中已标记缺陷样本的数量, $X_j^{S_i}$ 表示第 i 个源域中第 j 个缺陷样本, $Y_j^{S_i}$ 表示其对应的缺陷类别标签.令 $D^{T}=\left\{\left(X_j^{T}\right)\right\}_{j=1}^{n_T}$ ,其中 $n_T$ 表示目标域中未标记缺陷样本的数量, $X_j^{T}$ 表示目标域中第 j 个缺陷样本.此外, $\left\{\left(DL_j\right)\right\}_{j=1}^{n_{all}}$ 表示所有域中缺陷样本的域标签,其中 $n_{all}=\sum n_{S_i}+n_T$ .因此,COPILOT 框架的目标是以已标记源项目缺陷数据集 $D^{S_i}$ 作为训练集,通过引入多源

域适应技术来训练高精度的分类模型,并利用该模型预测未标记目标项目数据集 $D^{\mathrm{T}}$ 中样本的缺陷类别标签.以下小节将详细介绍框架中的各个模块.

### 2.2 域对抗训练模块

域对抗模块包含域编码器、梯度反转层(gradient reversal layer, GRL)[26]以及域鉴别器.该模块首先将所有软件项目的代码输入域编码器,以提取不同源域和目标域的特征表示向量.GRL 层位于域编码器和域鉴别器之间,通过在反向传播过程中实现梯度取反使得该层前后的网络训练目标相反,以实现对抗训练的目的.域鉴别器以经过梯度反转的特征表示向量作为输入,基于对抗训练的思想来进一步利用域不变特征,并最终输出不同源域与目标域之间的域相关性权重矩阵,以衡量不同源域和目标域之间的数据分布相似性.如式(1)所示,域鉴别器通过最小化训练损失目标 $L_{dc}$ 来学习不同源域和目标域之间的域不变表示,其中 n 表示当前训练轮次的缺陷样本数量, $E_d\left(X_j\right)$ 表示第 j 个缺陷样本 $X_j$ 经过域编码器 $E_d$ 后得到的特征表示, $DL_j$ 表示对应缺陷样本的域标签,D 表示域鉴别器,参数 $\lambda$ 会随着训练轮次的增加逐渐从 0 变化到 1.域鉴别器 D 的目标是根据域编码器 $E_d$ 生成的特征表示来判断缺陷样本所来源的域,而域编码器 $E_d$ 的目标则是生成的特征表示无法被域鉴别器 D 区分.通过 GRL 层进行梯度反转以进行对抗训练并达到收敛后,将混淆域鉴别器 D 预测缺陷样本的域标签的能力,同时输出对应的混淆后的概率矩阵 $p \in \mathbf{R}^{n\times(M+1)}$,其中 M 表示源域数量,1 则表示目标域.如式(2)所示,通过选取与源域相关的前 M 项作为域相关性矩阵 $P_d \in \mathbf{R}^{n\times M}$,并利用 softmax 函数计算出源域与目标域的域相关系数 $w$,其中 $w_i$ 表示第 i 个源域与目标域的相关性.该系数将作为权重影响后续模块中模型参数的学习过程.

$$L_{dc} = \lambda \frac{1}{n}\sum_{j=1}^{n} DL_j^{\mathrm{T}} \ln D\left(E_d\left(X_j\right)\right) \tag{1}$$

$$\left\{\left(w_i\right)\right\}_{i=1}^{M} = \mathrm{softmax}\left(P_d\right) \tag{2}$$

### 2.3 加权最大均值差异模块

在计算加权最大均值差异阶段,该模块首先使用与域编码器架构相同的特征编码器来提取不同源域和目标域的特征表示向量.为了进一步缓解不同源域与目标域之间的数据分布差异,该模块使用针对无监督域适应问题进行改进的域分布差异度量 WMMD[17].与传统的最大均值差异 MMD(maximum mean discrepancy)度量相比,WMMD 针对源域与目标域存在类别分布差异的情况进行加权处理.具体来说,给定 M 个源域 $\left\{D^{S_1}, D^{S_2}, \cdots, D^{S_M}\right\}$ 和目标域 $D^{\mathrm{T}}$,则第 i 个源域 $D^{S_i}$ 与目标域 $D^{\mathrm{T}}$ 的最大均值差异度量 MMD 可由式(3)计算,其中 $n_{S_i}$ 和 $n_T$ 分别表示源域 $D^{S_i}$ 与目标域 $D^{T}$ 中缺陷样本的数量, $H$ 表示用高斯核函数 $\phi(\cdot)$ 将缺陷样本映射到高维空间中进行度量.从式(3)中可以看出,计算 MMD 并不需要缺陷样本的标签信息,即 MMD 衡量的只是源域和目标域缺陷样本之间的边缘分布差异,近似于对齐源域和目标域的条件分布.在实际软件项目中,不同的源项目和目标项目之间的缺陷类别标签分布通常存在很大差异.而 MMD 度量则假设所有源域具有相同的权重,无法有效解决不同域中存在类别权重偏差的场景.因此本文使用改进的 WMMD 度量来缓解不同域在同一

缺陷类别标签上可能存在权重不等的问题.如 2.2 节所述,该模块中使用域相关性权重矩阵作为 WMMD 的加权策略.如式(4)所示,在计算 WMMD 时通过利用源域与目标域的相关性系数对源域进行加权,使得在训练分类器时可以更好地契合每个源域与目标域的数据分布,且能够自动减少域间类别权重偏差所带来的负面影响,从而最终达到最小化训练损失目标误差 $L_{dis}$ 的目的.总的来说,WMMD 模块通过利用域相关性可以减小目标域与高相关性源域之间的数据分布差异,从而将高相关性源域中的域私有特征更好地适配到目标域中.

$$MMD\left(D^{S_i}, D^{T}\right)=\left\|\frac{1}{n_{S_i}}\sum_{j=1}^{n_{S_i}}\phi\left(X_j^{S_i}\right)-\frac{1}{n_T}\sum_{k=1}^{n_T}\phi\left(X_k^{T}\right)\right\|_H^2 \tag{3}$$

$$L_{dis}=WMMD\left(D^{S}, D^{T}\right)=\sum_{i=1}^{M}w_i\left(MMD\left(D^{S_i}, D^{T}\right)\right) \tag{4}$$

### 2.4 缺陷类别预测模块

如图 2 所示,为了将训练好的模型应用于缺陷类别预测任务中,COPILOT 框架在该模块中使用的分类器为单层全连接层,将分类器的输出输入到 softmax 函数中得到样本对于每个缺陷类别的预测概率,最终选择概率最高的类别作为模型的预测结果.COPILOT 框架通过利用融合对抗训练(第 2.2 节)和注意力机制(第 2.3 节)的多源域适应技术来对齐不同源域与目标域的数据分布,但每个源域上的表示学习对于分类结果也是至关重要的.如式(5)所示,分类器通过最小化训练损失目标 $L_C$ 来缓解与目标域样本差异较大的源域样本导致的分类误差所引起负迁移问题,其中 M 表示源域的数量,$n_{S_i}$ 表示第 i 个源域中缺陷样本的数量,$E_f\left(X_j^{S_i}\right)$表示第 i 个源域中第 j 个缺陷样本经过特征编码器 $E_f$ 后得到的特征表示,$Y_j^{S_i}$ 表示对应的缺陷类别标签,C 表示分类器.

$$L_C=-\frac{1}{M}\sum_{i=1}^{M}\sum_{j=1}^{n_{S_i}}Y_j^{S_i\,\mathrm{T}}\ln C\left(E_f\left(X_j^{S_i}\right)\right) \tag{5}$$

综上,COPILOT 框架的整体损失函数可由式(6)计算.可分为两个部分更新参数,首先通过最小化对抗损失 $L_{dc}$ 来混淆域对抗器 D 的辨别能力.然后对分类误差 $L_C$ 进行反向传播,并对特征编码器 $E_f$ 和域编码器 $E_d$ 的参数进行了更新.在推理阶段,域对抗器 D 将被用作概率分布估计,并度量样本到域的关系.

$$LOSS=\min_{E_f,C}L_{dc}+\alpha\min_{E_f}L_{dis}+\min_{E_d,D}L_C \tag{6}$$

## 3 实验设置

### 3.1 实验数据集

本文构建的缺陷数据集的具体信息如表 1 所示.这些数据集均来自美国国家标准与技术研究院发起的 STONESOUP 计划,该计划旨在开发全面的自动化技术使最终用户能够安全地执行软件,并且能够让非专业人员降低发现和消除软件漏洞的时间及人力成本[27].具体来说,本文在数据集的选择上遵循以下标准.

- **流行性:** 所有项目均由 Java 语言编写,Java 是一种广泛使用的面向对象编程语言,在 TIOBE 编程语言排行榜中长期处于领先地位[28].此外,这八个项目在 GitHub 网站上获得的 Star 数均超过 100(见表 1 第 3 列),也证明这些项目在开源社区具备活跃性.
- **多样性:** 如表 1 第 2 列所示,这些项目涵盖了不同的功能领域(如测试工具、搜索软件、游戏引擎等),符合本文所提出的面向多源域适应技术的任务设定.
- **缺陷样本数量充足:** 这些项目的代码行数(lines of code, LoC)从 74k 到 542k 不等,包含足够的缺陷样本用于模型训练.表 1 第 5 列展示了每个项目中包含的缺陷数量,范围从 153 到 437.每个项目的缺陷样本中都包含常见的 44 种 CWE 缺陷类别(如表 2 所示).表 1 最后 4 列按缺陷数量递减的顺序展示了

严重性排名在 CWE Top 25 的缺陷类别,它们的缺陷风险得分依次为 CWE-89(最新得分 22.11)、CWE-190(最新得分 6.53)、CWE-400(最新得分 3.56)以及 CWE-78(最新得分 17.53).该排名通过将 CWE 成为漏洞根源的频率与 CVSS(common vulnerability scoring system)所衡量的每个漏洞利用的平均严重性结合在一起进行计算排序[29].

- **缺陷样本标签可获取:** 每个样本数据所对应的都是开源项目软件代码中一段带缺陷的 Java 函数,因此本文使用 STONESOUP 计划中收集的 CWE 缺陷类别结果作为数据集的标签以保证可靠性.

**表 1** 实验数据集的具体信息

| 项目 | 功能领域 | GitHub Star 数 | LoC | 缺陷数量 | 严重缺陷类别及其数量比例 | | | |
|---|---|---|---|---|---|---|---|---|
| | | | | | CWE-89 | CWE-190 | CWE-400 | CWE-78 |
| Apache JMeter | 负载测试工具 | 7.3k | 123k | 153 | 12 (7.84%) | 6 (3.92%) | 6 (3.92%) | 3 (1.96%) |
| Apache Jena | 语义网应用框架 | 985 | 413k | 426 | 39 (9.15%) | 15 (3.52%) | 14 (3.29%) | 4 (0.94%) |
| Apache Lenya | 内容管理工具 | 170 | 433k | 390 | 42 (10.77%) | 12 (3.08%) | 11 (2.82%) | 5 (1.28%) |
| Apache Lucene | 搜索软件 | 1.9k | 450k | 422 | 38 (9.00%) | 14 (3.32%) | 12 (2.84%) | 2 (0.47%) |
| Apache POI | 文件读写库 | 1.7k | 337k | 437 | 40 (9.15%) | 16 (3.66%) | 13 (2.97%) | 4 (0.92%) |
| CoffeeMUD | 游戏引擎 | 167 | 542k | 428 | 41 (9.58%) | 19 (4.44%) | 12 (2.80%) | 2 (0.47%) |
| Elasticsearch | 分布式搜索引擎 | 65k | 367k | 430 | 39 (9.07%) | 20 (4.65%) | 13 (3.02%) | 5 (1.16%) |
| JTree | 树形符号 SDK | 373 | 74k | 153 | 12 (7.84%) | 6 (3.92%) | 6 (3.92%) | 3 (1.96%) |

**表 2** CWE 缺陷类别分类

| 类型 | CWE 缺陷类别 |
|---|---|
| 输入验证和数据完整性问题 | CWE-36, CWE-390, CWE-391, CWE-459, CWE-789 |
| 计算和代码注入问题 | CWE-78, CWE-363, CWE-543, CWE-839 |
| 身份验证和授权问题 | CWE-41, CWE-89, CWE-209, CWE-252, CWE-400, CWE-584, CWE-834, CWE-835 |
| 安全配置和管理问题 | CWE-23, CWE-190, CWE-412, CWE-832 |
| 缓冲区问题 | CWE-191, CWE-195, CWE-196, CWE-197, CWE-367, CWE-674, CWE-764, CWE-765, CWE-820, CWE-821, CWE-833 |
| Miscellaneous | CWE-88, CWE-774, CWE-194, CWE-253, CWE-369, CWE-414, CWE-460, CWE-564, CWE-567, CWE-606, CWE-609, CWE-663 |

## 3.2 实验对比方法

为了评估 COPILOT 框架在缺陷类别预测任务中的性能,本文将所提出的方法与以下 5 个与本文相关的基线方法进行比较.

- μVulDeePecker[7]: 一种基于深度学习的多类别漏洞检测方法,通过使用代码关注度的表示方法捕捉精确定位漏洞类别的信息,从而进行项目内的漏洞检测任务.
- AC-GAN[12]: 一种采用对抗学习思想的单源域适应方法,通过使用生成式对抗网络来减小目标项目与源项目之间的特征分布差异,从而实现跨项目软件缺陷预测.
- MDAN[19]: 一种应用于计算机视觉领域的多源域对抗神经网络模型,通过平滑逼近优化提出了新的泛化界,从而使数据和任务适应模型更加有效.
- M3SDA[20]: 一种基于矩匹配的多源域适应方法,核心思想是通过学习一个共享的表示空间,将不同域和模态的数据融合起来,并且通过一个多源域自适应的框架来进行知识迁移.
- ABMSDA[21]: 一种基于注意力机制的多源域适应方法,根据域相关性提出 WMD 以更多地关注相似性较高的源域,通过将多个源域的数据进行注意力加权来提高模型的泛化能力.

## 3.3 研究问题

为了评估基于多源域适应的缺陷类别预测方法 COPILOT 的有效性,本文研究了以下三个研究问题(RQ, research question).

- **RQ1(总体实验):** COPILOT 在缺陷类别预测任务中能否比基线方法获得更好的模型性能?
- **RQ2(消融实验):** COPILOT 框架中的每个模块对缺陷类别预测任务是否有效?
- **RQ3(不同场景设定下的分析实验):** COPILOT 在处理不同类型的缺陷、严重类型的缺陷以及不同数据量的缺陷时能否比基线方法获得更好的模型性能?

### 3.4 实验评估指标

由于不同实验场景设定的差异,合理选择评估指标来衡量模型的分类结果是有必要的[30].本文针对不同的研究问题采用不同的度量指标来评估缺陷类别预测模型的预测性能.总的来说,分类任务的实验结果可以由混淆矩阵表示,它显示了模型预测结果和人工标注标签之间的差异.对于缺陷类别预测场景来说,混淆矩阵中的实例可划分为四类: 1)真阳性实例(true positive, TP): 将真实缺陷类别正确预测为真实缺陷类别; 2)假阳性实例(false positive, FP): 将其他缺陷类别错误预测为真实缺陷类别; 3)真阴性实例(true negative, TN): 将其他缺陷类别正确预测为其他缺陷类别; 4)假阴性实例(false negative, FN): 将真实缺陷类别错误预测为其他缺陷类别.

RQ1 和 RQ2 的实验场景针对模型的整体性能进行分析,即统计模型在所有 44 类缺陷类别的预测结果.因此,本文选择了以下三个在已有缺陷相关研究[31-33]中被广泛采用的综合度量指标来评估实验结果.

- 准确率(accuracy, Acc): 准确率是分类问题中最直观的评价指标,是一个针对所有样本的统计量.准确率是指分类正确的样本占总样本个数的比例,其计算方法如式(7)所示.准确率的值介于 0 到 1 之间,值越高表示模型的预测性能越好.

$$Acc = \frac{TP + TN}{TP + FP + TN + FN} \tag{7}$$

- 马修斯相关系数(Matthews correlation coefficient, MCC): MCC 指标同时考虑混淆矩阵中所有实例类别的值,因此 MCC 通常被视为在分类任务中描述混淆矩阵最具信息性的系数之一.如式(8)所示,MCC 返回介于-1 到 1 之间的值.系数的值越接近 1 表示预测和观察之间的一致性越高,即模型性能越好.

$$MCC = \frac{TP \times TN - FP \times FN}{\sqrt{(TP + FP)\times(TP + FN)\times(TN + FP)\times(TN + FN)}} \tag{8}$$

- Kappa 系数(Kappa): 该系数是一种衡量分类精度的一致性检验指标,系数的值越高则代表模型实现的分类准确度越高.Kappa 系数可由式(9)和(10)计算而得,其值的取值范围是[0,1].其中 Acc 是分类器之间的观察一致性,Q 是分类器之间的随机一致性,可以通过计算分类器在数据集上的预期准确率而得.Kappa 系数的值大于 0.6 则认为预测和观察之间具有高度的一致性.

$$Kappa = \frac{Acc - Q}{1 - Q} \tag{9}$$

$$Q = \frac{(TP + FP)\times(TP + FN)\times(TN + FP)\times(TN + FN)}{(TP + FP + TN + FN)^2} \tag{10}$$

RQ3 的实验场景则需要分析模型对不同类型缺陷的分类结果.因此,本文选择适用于分析单一类别评估的指标 F1 值,它是精准率(precision, P)和召回率(recall, R)的加权调和平均.令缺陷类别 $CWE_i$ 在混淆矩阵中的真阳性实例,假阳性实例,真阴性实例以及假阴性实例分别为 $TP_i$ , $FP_i$ , $TN_i$ 以及 $FN_i$ ,则 $CWE_i$ 的 F1 值可通过式(13)计算.此外,针对表 1 中列出的 4 个严重缺陷类别,本文通过结合 3.1 节中描述的缺陷风险得分[29]对 F1 评估指标进行加权处理.如式(14)-(16)所示,其中 $w_\{*\}$ 表示加权后的评估指标,score 表示缺陷风险得分.

$$P_i = \frac{TP_i}{TP_i + FP_i} \tag{11}$$

$$R_i = \frac{TP_i}{TP_i + FN_i} \tag{12}$$

$$F1_i = \frac{2 \times P_i \times R_i}{P_i + R_i} \tag{13}$$

$$w_P = \frac{\text{score}(CWE_i)}{\sum_{j=1}^{4} \text{score}(CWE_i)} \times P_i \tag{14}$$

$$w_R = \frac{\text{score}(CWE_i)}{\sum_{j=1}^{4}\text{score}(CWE_i)} \times R_i \tag{15}$$

$$w_F1 = \frac{\text{score}(CWE_i)}{\sum_{j=1}^{4}\text{score}(CWE_i)} \times F1_i \tag{16}$$

### 3.5 实验设计

针对跨项目场景下的缺陷类别预测任务,本文按顺序从八个项目中选取一个项目作为目标项目(目标域),然后利用其余项目作为源项目(源域),因此总共有八种可能的项目(域)组合.对于每个域组合,源域中的七个项目作为训练集用于 COPILOT 框架中模型的参数学习过程,然后利用经过多源域适应技术选择的源域历史缺陷数据训练的 softmax 分类器对目标域中缺陷样本的类别进行预测.

本文利用作者论文中开源的代码对基线模型 AC-GAN、MDAN、M3SDA 以及 ABMSDA 进行复现.针对 μVulDeePecker 这个基线方法,本文通过已有论文[7]中描述的相同的网络结构以及参数来复现相关模型.本文所提出的 COPILOT 框架利用 CodeT5[22]作为基础模型架构用于编码特征提取(即域编码器和特征编码器),CodeT5 是由 6 层的 Transformer 编码器-解码器架构[23]组成.此外,域鉴别器为输出大小为 8(即域数量)的全连接层,分类器为输出大小为 44(即缺陷类别数量)的全连接层.在 COPILOT 框架中,源和目标项目的最大输入序列长度为 800.公式(6)中的超参数$\alpha$设置为 0.01.在训练策略中,COPILOT 采用 AdamW 优化器,学习率设置为 5e-5,批处理大小为 8,训练的 epoch 为 30.在每个训练 epoch 中,模型将目标域的缺陷样本以 2:1 的比例划分,其中三分之一的数据作为验证集,三分之二的数据作为测试集.若模型在验证集上的评估指标在新一轮训练中得到改善,则保存相关模型及其参数.本文采用早停(early stopping)机制以防止模型过拟合,即当模型在验证集上经过 2 个 epoch 仍没有效果提升时,训练将提前结束.在测试阶段,使用训练好的模型在测试集进行测试,并根据 3.4 节中提出的评估指标进行比较实验.本文中的所有实验均在 PyTorch 环境下实现,并在装有 32G 内存的 Iluvatar BI-V100 GPU 的服务器上进行(代码已发布在 https://github.com/starve123456/copilot).

对于 3.4 节提出的评估指标,本文采用 Scott-Knott Effect Size Difference(ESD)统计检验方法对 COPILOT 框架与基线方法进行排名.Scott-Knott ESD 是一种均值比较方法,它利用层次聚类将数据集划分为统计学上差异不可忽略的不同的组[34].本文进一步采用 Cohen 效应量 d 来评估 COPILOT 框架与对比模型之间效果的差异的重要性.当$|d| \le 0.2$时,模型间的差异可忽略(N, negligible);当$0.2 < |d| \le 0.5$时,模型间的差异小(S, small);当$0.5 < |d| \le 0.8$时,模型间的差异是中等(M, medium);当$|d| \ge 0.8$时,模型间的差异大(L, large).当模型 A 对模型 B 的效应量 d 为正且大于 0.2,则表明模型 A 具有更高的使用价值.

## 4 实验结果与分析

### 4.1 RQ1实验结果与分析

表 3-表 5 分别列出了 COPILOT 和五个基线方法在每个目标项目上取得的评估指标结果(每行代表一个目标项目对应的结果),并加粗表示最好的结果.对于每个方法,Avg.行是其在所有目标项目上取得结果的平均值.Cohen's $d$ 行表示 COPILOT 与每个基线方法之间的 Cohen 效应量值.图 3 对 Scott-Knott ESD 检验的排名结果进行了可视化展示.据此,可以得出下述分析结果.

- COPILOT 在所有评估指标的 Scott-Knott ESD 检验排名中都取得了最好的结果,说明 COPILOT 的性能对比基线方法具有一定优势.具体地,与项目内多分类漏洞检测方法μVulDeePecker 以及基于单源域适应的缺陷预测方法 AC-GAN 相比,COPILOT 充分利用来自不同源项目中的代码语义信息,并利用多源域适应策略来减小不同项目之间的数据分布差异,因此取得了优于基于单源域适应方法的模型预测性能.此外,其余三个应用于计算机视觉领域的多源域适应模型(即 MDAN、M3SDA 以及

ABMSDA)缺乏对代码信息的关注,无法有效捕获代码语义信息,因此在针对软件工程领域的缺陷类别预测任务上的模型性能不及 COPILOT.

- 针对 Acc 来说,COPILOT 在 8 个目标项目上取得的平均 Acc 为 0.947.与μVulDeePecker、AC-GAN、MDAN、M3SDA 和 ABMSDA 相比分别提高了 23.6%、15.5%、2.2%、2.1%和 26.2%.针对 MCC 来说,COPILOT 在 8 个目标项目上取得的平均 MCC 为 0.946.与μVulDeePecker、AC-GAN、MDAN、M3SDA 和 ABMSDA 相比分别提高了 24.2%、15.9%、2.2%、2.0%和 27.7%.针对 Kappa 来说,COPILOT 在 8 个目标项目上取得的平均为 0.945.与μVulDeePecker、AC-GAN、MDAN、M3SDA 和 ABMSDA 相比分别提高了 24.7%、16.1%、2.2%、2.1%和 27.5%.
- COPILOT 在所有评估指标中相对于μVulDeePecker、AC-GAN、MDAN、M3SDA 和 ABMSDA 的 Cohen 效应量值衡量的差异大小分别为 L、L、M、M 和 L,在统计意义上展示了 COPILOT 对比基线方法的性能优势,也充分表明跨项目场景下 COPILOT 框架的实用性.
- 特别地,COPILOT 在使用 Apache POI 作为目标域时的评估指标实验结果比对比方法 M3SDA 略低,但在整体模型性能上具有中等大小的差异优势.通过分析可以发现 Apache POI 项目的功能是提供一组复杂的 API 来操作 Microsoft Office 格式的文件.这些 API 具有复杂的互依赖关系,并且涉及的代码量庞大.因此预测该类代码中的缺陷类别是一个挑战,将作为未来的研究工作继续进行.

**表 3** RQ1 对比实验结果(Acc)

| 目标项目 | μVulDeePecker | AC-GAN | MDAN | M3SDA | ABMSDA | COPILOT |
|---|---|---|---|---|---|---|
| Apache JMeter | 0.726 | 0.794 | 0.971 | 0.922 | 0.500 | **1.000** |
| Apache Jena | 0.671 | 0.821 | 0.907 | 0.880 | 0.764 | **0.914** |
| Apache Lenya | 0.795 | 0.890 | 0.897 | 0.908 | 0.901 | **0.930** |
| Apache Lucene | 0.828 | 0.731 | 0.936 | **0.943** | 0.879 | **0.943** |
| Apache POI | 0.632 | 0.866 | 0.925 | **0.945** | 0.889 | 0.941 |
| CoffeeMUD | 0.899 | 0.896 | 0.903 | **0.920** | 0.826 | **0.920** |
| Elasticsearch | 0.765 | 0.913 | 0.899 | 0.940 | 0.876 | **0.950** |
| JTree | 0.816 | 0.651 | **0.981** | 0.971 | 0.369 | **0.981** |
| **Avg.** | 0.766 | 0.820 | 0.927 | 0.928 | 0.750 | **0.947** |
| **Cohen's *d*** | 2.770 (L) | 1.934 (L) | 0.638 (M) | 0.658 (M) | 1.357 (L) | - |

**表 4** RQ1 对比实验结果(MCC)

| 目标项目 | μVulDeePecker | AC-GAN | MDAN | M3SDA | ABMSDA | COPILOT |
|---|---|---|---|---|---|---|
| Apache JMeter | 0.723 | 0.792 | 0.970 | 0.920 | 0.460 | **1.000** |
| Apache Jena | 0.667 | 0.816 | 0.904 | 0.879 | 0.758 | **0.911** |
| Apache Lenya | 0.790 | 0.887 | 0.894 | 0.906 | 0.898 | **0.928** |
| Apache Lucene | 0.825 | 0.724 | 0.934 | **0.941** | 0.875 | **0.941** |
| Apache POI | 0.621 | 0.863 | 0.923 | **0.943** | 0.886 | 0.940 |
| CoffeeMUD | 0.896 | 0.893 | 0.900 | **0.917** | 0.820 | **0.917** |
| Elasticsearch | 0.758 | 0.910 | 0.896 | 0.938 | 0.872 | **0.948** |
| JTree | 0.812 | 0.645 | **0.980** | 0.970 | 0.353 | **0.980** |
| **Avg.** | 0.761 | 0.816 | 0.925 | 0.927 | 0.740 | **0.946** |
| **Cohen's *d*** | 2.765 (L) | 1.941 (L) | 0.638 (M) | 0.652 (M) | 1.351 (L) | - |

**表 5** RQ1 对比实验结果(Kappa)

| 目标项目 | μVulDeePecker | AC-GAN | MDAN | M3SDA | ABMSDA | COPILOT |
|---|---|---|---|---|---|---|
| Apache JMeter | 0.717 | 0.788 | 0.970 | 0.919 | 0.481 | **1.000** |
| Apache Jena | 0.659 | 0.814 | 0.904 | 0.876 | 0.756 | **0.911** |
| Apache Lenya | 0.788 | 0.886 | 0.894 | 0.905 | 0.897 | **0.928** |
| Apache Lucene | 0.822 | 0.720 | 0.934 | **0.941** | 0.875 | **0.941** |
| Apache POI | 0.619 | 0.862 | 0.923 | **0.943** | 0.885 | 0.939 |
| CoffeeMUD | 0.896 | 0.892 | 0.899 | **0.917** | 0.819 | **0.917** |
| Elasticsearch | 0.756 | 0.910 | 0.896 | 0.937 | 0.872 | **0.948** |
| JTree | 0.810 | 0.640 | **0.980** | 0.970 | 0.350 | **0.980** |
| **Avg.** | 0.758 | 0.814 | 0.925 | 0.926 | 0.742 | **0.945** |
| **Cohen's *d*** | 2.762 (L) | 1.934 (L) | 0.634 (M) | 0.657 (M) | 1.361 (L) | - |

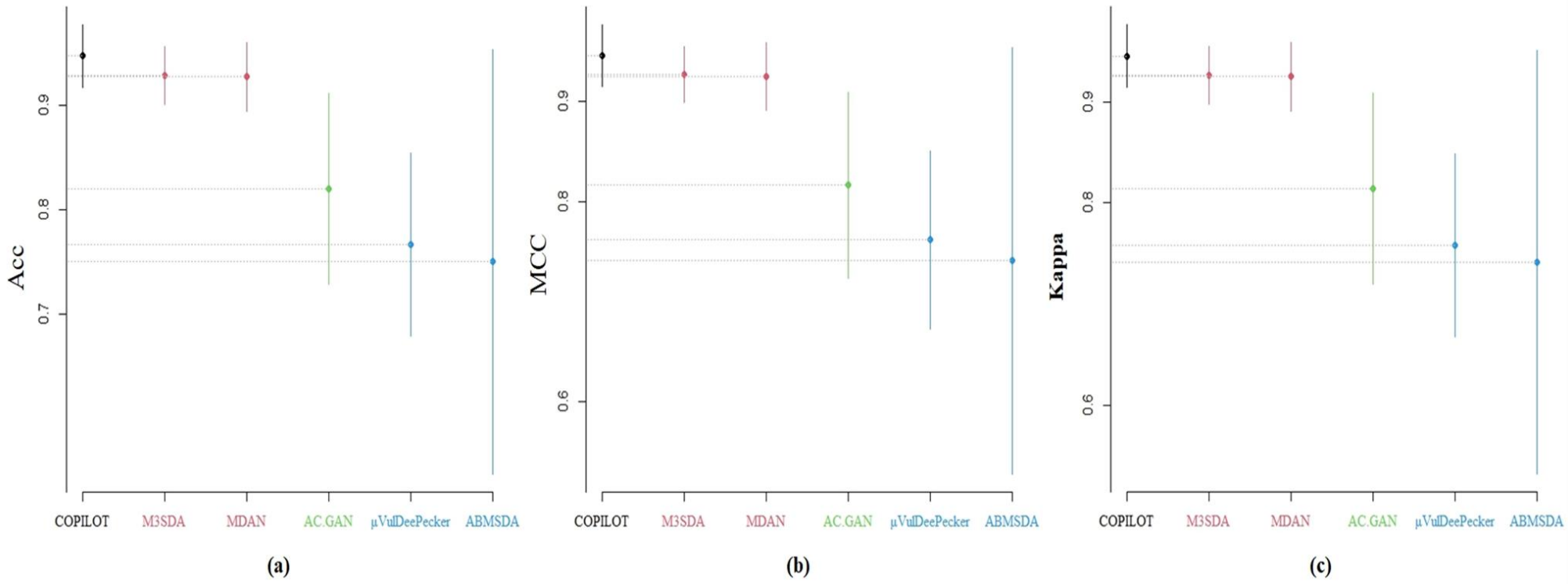

图 3　RQ1 的 Scott-Knott ESD 结果可视化. (a) Acc; (b) MCC; (c) Kappa

### 4.2　　RQ2实验结果与分析

本节的消融实验通过将 COPILOT 框架中的域对抗训练(adversarial training, AT)模块以及加权最大均值差异(WMMD)模块分别剔除,通过跨项目缺陷类别预测任务进行比较.其中,$\text{COPILOT}_{\text{w/o AT}}$是仅基于加权最大均值差异模块的模型,$\text{COPILOT}_{\text{w/o WMMD}}$是仅基于域对抗训练模块的模型.为了进行公平的实验,本文采取与 3.5 节相同的设置训练 $\text{COPILOT}_{\text{w/o AT}}$ 与 $\text{COPILOT}_{\text{w/o WMMD}}$.

表 6-表 8 分别列出了 COPILOT 分别与 $\text{COPILOT}_{\text{w/o AT}}$ 和 $\text{COPILOT}_{\text{w/o WMMD}}$ 在每个目标项目上取得的评估指标结果,并加粗表示最好的结果.可以看到,在所有三个指标上,COPILOT 在 8 个目标项目中(除 CoffeeMUD 外)均展现出了最佳效果.如图 4 所示,在与仅使用单一模块进行训练的基线方法相比时,本文所提出的 COPILOT 框架在三个评估指标中的 Scott-Knott ESD 统计检验排名都比两个基线方法要高.具体来说,与基线方法 $\text{COPILOT}_{\text{w/o AT}}$ 和 $\text{COPILOT}_{\text{w/o WMMD}}$ 相比,本文所提出的方法 COPILOT 在平均 Acc 值的相对提升幅度分别为 1.9%和 1.1%,在平均 Kappa 值的相对提升幅度分别为 1.9%和 1.1%.针对 Cohen 效应量结果,COPILOT 相对于 $\text{COPILOT}_{\text{w/o AT}}$ 在所有评估指标下有中等幅度的优势,而 COPILOT 相对于 $\text{COPILOT}_{\text{w/o WMMD}}$ 在三个评估指标下有小幅度的优势,因此模型的性能差异也具有统计显著性.针对不同目标项目进行分析可以看到,对于 Apache JMeter 和 JTree 等数据量较小的项目来说,在不使用 WMMD 模块的情况下,只进行对抗训练就能获得最佳结果.而对于 CoffeeMUD 项目来说,WMMD 模块则对模型导致负面的效果.总的来说,COPILOT 通过融合对抗训练以及使用域相关性作为注意力机制学习到更加通用的代码语义表示以减小不同项目间的数据差异,从而在跨项目场景下提高模型在缺陷类别预测任务中的性能.

**表 6**　RQ2 消融实验结果(Acc)

| 目标项目 | $\text{COPILOT}_{\text{w/o AT}}$ | $\text{COPILOT}_{\text{w/o WMMD}}$ | COPILOT |
|---|---|---|---|
| Apache JMeter | **1.000** | **1.000** | **1.000** |
| Apache Jena | 0.897 | 0.907 | **0.914** |
| Apache Lenya | 0.908 | 0.905 | **0.930** |
| Apache Lucene | 0.923 | 0.926 | **0.943** |
| Apache POI | 0.938 | 0.932 | **0.941** |
| CoffeeMUD | 0.916 | **0.923** | 0.920 |
| Elasticsearch | 0.933 | 0.926 | **0.950** |
| JTree | 0.922 | **0.981** | **0.981** |
| **Avg.** | 0.930 | 0.937 | **0.947** |
| **Cohen's *d*** | 0.583 (M) | 0.308 (S) | - |

**表 7**　RQ2 消融实验结果(MCC)

| 目标项目 | $\text{COPILOT}_{\text{w/o AT}}$ | $\text{COPILOT}_{\text{w/o WMMD}}$ | COPILOT |
|---|---|---|---|
| Apache JMeter | **1.000** | **1.000** | **1.000** |
| Apache Jena | 0.895 | 0.905 | **0.911** |
| Apache Lenya | 0.908 | 0.902 | **0.928** |
| Apache Lucene | 0.920 | 0.924 | **0.941** |
| Apache POI | 0.936 | 0.930 | **0.940** |
| CoffeeMUD | 0.914 | **0.921** | 0.917 |
| Elasticsearch | 0.931 | 0.924 | **0.948** |
| JTree | 0.921 | **0.980** | **0.980** |
| **Avg.** | 0.928 | 0.936 | **0.946** |
| **Cohen's *d*** | 0.562 (M) | 0.299 (S) | - |

**表 8** RQ2 消融实验结果(Kappa)

| 目标项目 | $\textbf{COPILOT}_{w/o\ AT}$ | $\textbf{COPILOT}_{w/o\ WMMD}$ | **COPILOT** |
|---|---|---|---|
| Apache JMeter | **1.000** | **1.000** | **1.000** |
| Apache Jena | 0.893 | 0.904 | **0.911** |
| Apache Lenya | 0.905 | 0.901 | **0.928** |
| Apache Lucene | 0.920 | 0.923 | **0.941** |
| Apache POI | 0.936 | 0.929 | **0.939** |
| CoffeeMUD | 0.913 | **0.920** | 0.917 |
| Elasticsearch | 0.931 | 0.924 | **0.948** |
| JTree | 0.920 | **0.980** | **0.980** |
| **Avg.** | 0.927 | 0.935 | **0.945** |
| **Cohen's *d*** | 0.579 (M) | 0.313 (S) | - |

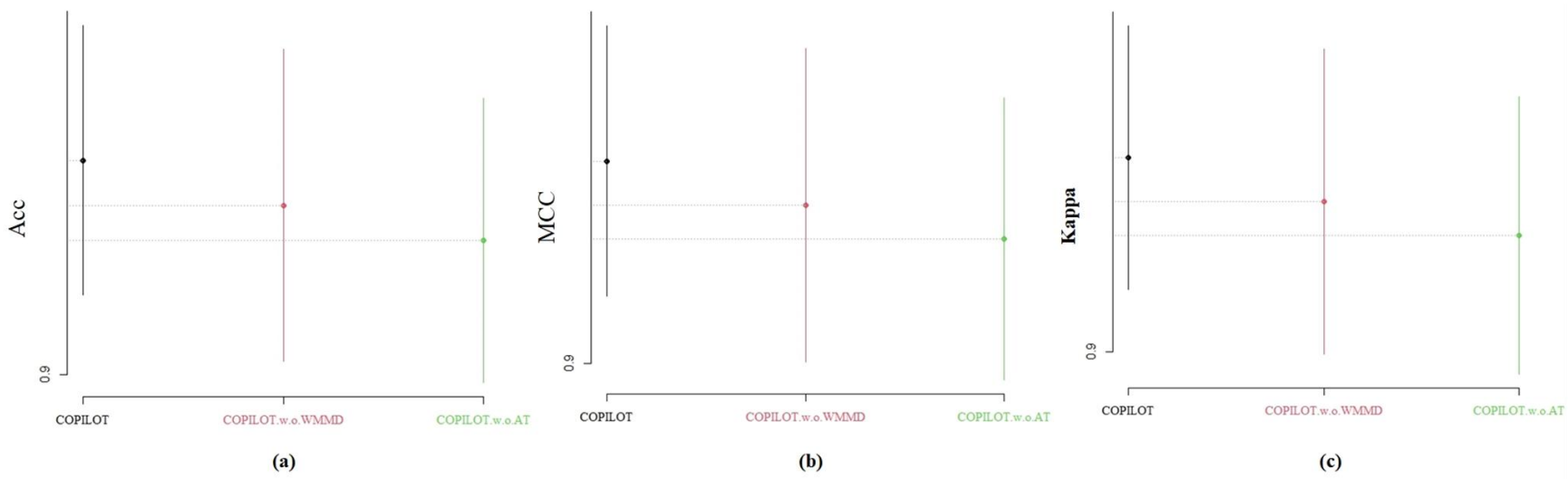

图 4 RQ2 的 Scott-Knott ESD 结果可视化. (a) Acc; (b) MCC; (c) Kappa

### 4.3 RQ3实验结果与分析

本节通过对不同类型的缺陷、严重类型的缺陷以及不同数据量的缺陷的类型进行分析,对 COPILOT 的缺陷类别预测性能进行更充分的验证.

#### 4.3.1 不同类型的缺陷分析

如表 2 所示,本文将 44 种 CWE 缺陷类别分为六类,即输入验证和数据完整性问题(Type1)、计算和代码注入问题(Type2)、身份验证和授权问题(Type3)、安全配置和管理问题(Type4)、缓冲区问题(Type5)以及 Miscellaneous(Type6).表 9 列出了 COPILOT 和五个基线方法在 Type1-Type6 上取得的 F1 结果,并加粗表示最好的结果.具体来说, COPILOT 在 6 种类型中(除 Type1 外)均展现出了最佳效果.针对 F1 来说,COPILOT 在 6 种类型上取得的平均 F1 为 0.932.与μVulDeePecker、AC-GAN、MDAN、M3SDA 和 ABMSDA 相比分别提高了 27.4%、25.0%、3.2%、3.3%和 36.4%.图 5 展示了在不同类型缺陷场景中的 Scott-Knott ESD 检验排名结果,COPILOT 排名第一,说明利用 COPILOT 方法所构建的缺陷类别预测模型并不是只能预测特定类型的缺陷类别,在所有类型的缺陷中对比其它基线方法都能够具有显著的性能提升.同时,COPILOT 在 F1 指标下相对于μVulDeePecker、AC-GAN、MDAN、M3SDA 和 ABMSDA 的 Cohen 效应量 d 分别为 3.205(L)、3.754(L)、0.594(M)、0.484(S)和 2.997(L),在统计意义上验证了 COPILOT 的优异性能.

**表 9** RQ3 对比实验结果(F1)

| 类型 | **μVulDeePecker** | **AC-GAN** | **MDAN** | **M3SDA** | **ABMSDA** | **COPILOT** |
|---|---|---|---|---|---|---|
| Type1 | 0.737 | 0.689 | 0.915 | **0.930** | 0.703 | 0.923 |
| Type2 | 0.579 | 0.716 | 0.794 | 0.747 | 0.464 | **0.853** |
| Type3 | 0.735 | 0.811 | 0.947 | 0.942 | 0.741 | **0.954** |
| Type4 | 0.784 | 0.768 | 0.906 | 0.938 | 0.721 | **0.952** |
| Type5 | 0.783 | 0.745 | 0.942 | 0.928 | 0.763 | **0.970** |
| Type6 | 0.773 | 0.816 | 0.915 | 0.931 | 0.709 | **0.943** |
| **Avg.** | 0.732 | 0.746 | 0.903 | 0.903 | 0.683 | **0.932** |
| **Cohen's *d*** | 3.205 (L) | 3.754 (L) | 0.594 (M) | 0.484 (S) | 2.997 (L) | - |

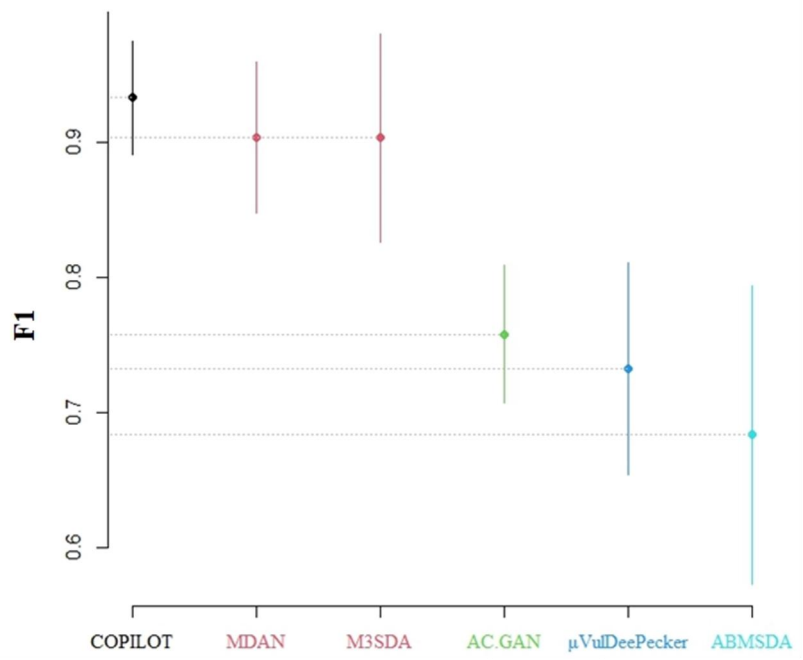

图 5　RQ3 中第一个分析实验的 Scott-Knott ESD 结果可视化

#### 4.3.2　严重类型的缺陷分析

本节分析了 COPILOT 以及基线方法在 4 类严重缺陷(如表 1 所示)类别场景下的对比实验.表 10 列出了 COPILOT 和五个基线方法在每个目标项目的严重缺陷类型上取得的 w_F1 结果,并加粗表示最好的结果.具体来说,在八个目标项目中,COPILOT 在 w_F1 指标下取得了五个项目中的最佳的结果.针对 w_F1 来说,COPILOT 在 8 个目标项目上取得的平均 w_F1 为 0.877.与μVulDeePecker、AC-GAN、MDAN、M3SDA 和 ABMSDA 相比分别提高了 23.2%、6.4%、9.2%、10.1%和 44.9%.如图 6 所示,COPILOT 在 *w*_F1 评估指标下的 Scott-Knott ESD 检验结果排名第一,说明该分析实验的结果具有统计显著性.同时,COPILOT 在 w_F1 指标下相对于μVulDeePecker、AC-GAN、MDAN、M3SDA 和 ABMSDA 的 Cohen 效应量 d 分别为 1.110(L)、0.356(S)、0.537(M)、0.503(M)和 1.456(L),表明 COPILOT 至少有小幅度的优势.综上,使用 COPILOT 方法在预测严重类型缺陷时也能够获得显著的性能提升.

**表 10**　RQ3 对比实验结果(*w*_F1)

| 目标项目 | μVulDeePecker | AC-GAN | MDAN | M3SDA | ABMSDA | COPILOT |
|---|---|---|---|---|---|---|
| Apache JMeter | 0.498 | 0.956 | 0.856 | 0.915 | 0.233 | **1.000** |
| Apache Jena | 0.567 | 0.618 | 0.788 | 0.491 | 0.533 | **0.802** |
| Apache Lenya | **0.877** | **0.877** | 0.690 | 0.749 | 0.692 | 0.769 |
| Apache Lucene | **0.941** | 0.846 | 0.591 | 0.593 | 0.588 | 0.597 |
| Apache POI | 0.643 | 0.829 | 0.725 | 0.891 | 0.834 | **0.917** |
| CoffeeMUD | 0.635 | **0.963** | 0.949 | 0.931 | 0.684 | 0.957 |
| Elasticsearch | 0.723 | 0.937 | 0.838 | 0.816 | 0.895 | **0.971** |
| JTree | 0.809 | 0.568 | 0.985 | 0.985 | 0.384 | **1.000** |
| **Avg.** | 0.712 | 0.824 | 0.803 | 0.796 | 0.605 | **0.877** |
| **Cohen's *d*** | 1.110 (L) | 0.356 (S) | 0.537 (M) | 0.503 (M) | 1.456 (L) | - |

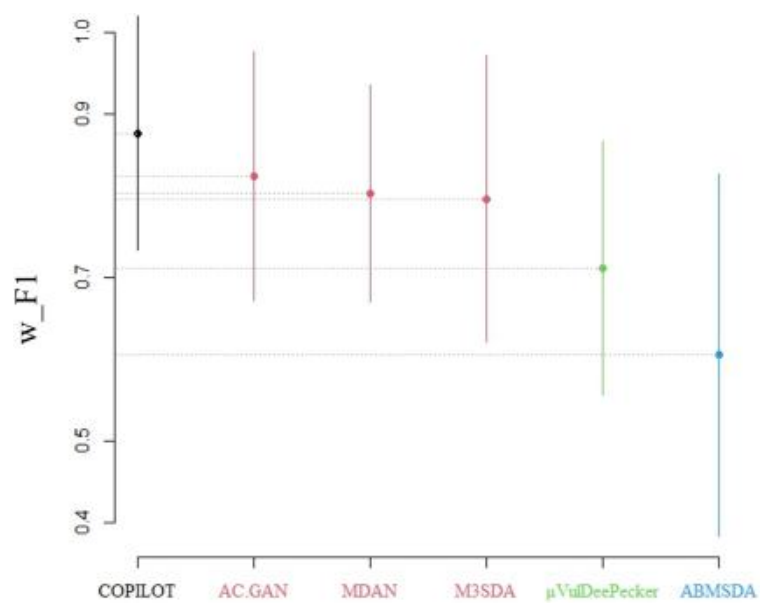

图 6 RQ3 中第二个分析实验的 Scott-Knott ESD 结果可视化

#### 4.3.3 不同数据量的缺陷分析

本节分析了 COPILOT 和五个基线方法在不同数据量缺陷场景下的对比实验.为了展示简洁,本节使用折线图来表示对比实验结果.如图 7 所示,横轴表示本文所包含的 44 种 CWE 缺陷类别,从左往右按数据量递减的顺序排列(如图 7 中灰色蓝点折线所示).左侧纵轴表示在本节所述场景下预测模型的 F1 评估指标值,右侧纵轴表示不同缺陷类别的样本数据量.总的来说,与基线方法相比,本文所提出的方法 COPILOT(如图 7 中黑色绿点折线所示)在几乎所有数据量大小的情况下都能获得最佳的结果.特别地,以缺陷类别 CWE-367 为界,当缺陷类别的样本数据量大小超过 36 时,COPILOT 方法能够比其他基线方法取得更稳定的模型性能.当缺陷类别的样本数据量小于 36 时,COPILOT 和五个基线方法的在这些类别的缺陷(即 CWE-78、CWE-88 以及 CWE-839)上性能都有所下降,但 COPILOT 的模型预测性能依旧能够超过其他基线方法.通过分析缺陷类别的特征可以发现,CWE-78 是一个与操作系统命令注入有关的缺陷,而 CWE-839 则涉及数值范围的比较,需要检测数值是否在特定范围内.这些缺陷出现的原因可能是由于跨系统命令使用导致的转义处理不恰当,或者需要涉及多个数值范围的比较.这些复杂性较高的缺陷类别影响模型的预测性能.因此,如何提高缺陷类别预测模型对缺陷语义的理解能力从而提高性能将是未来考虑的工作之一.图 8 则展示了在不同数据量下的 Scott-Knott ESD 检验结果,COPILOT 在 F1 评估指标下排名第一.综上,COPILOT 在不同数据量缺陷场景中也能取得最好的模型预测结果.

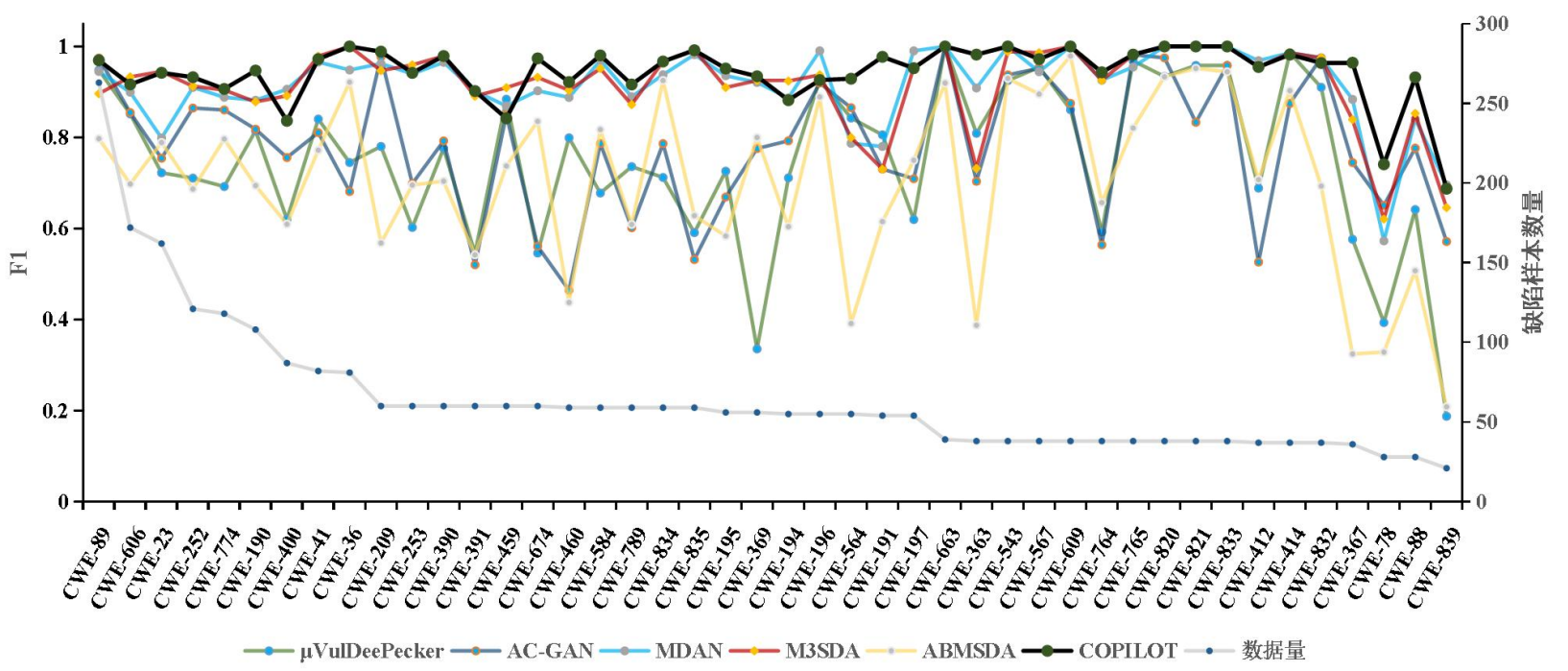

图 7 COPILOT 与其他基线方法在不同数据量缺陷场景中的 F1 指标比较结果

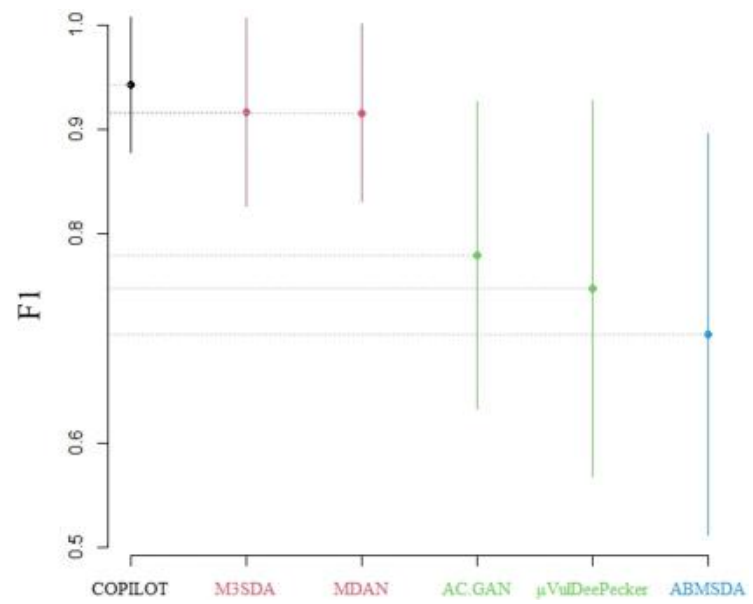

图 8 RQ3 中第三个分析实验的 Scott-Knott ESD 结果可视化

### 4.4 有效性威胁分析

以下分别从外部有效性、内部有效性和结构有效性对本文提出的 COPILOT 进行分析.

- 外部有效性: 本文所提方法 COPILOT 面临的主要外部有效性威胁是实验采用的数据集.本文涉及的研究所使用的数据集是从开源软件项目中挑选的八个真实缺陷项目.因此,当 COPILOT 在其他项目上进行训练时可能会导致更好或者更差的结果.在未来的工作中,我们将致力于构建更具有普遍性的缺陷类别数据库.
- 内部有效性: 内部有效性威胁来自模型训练时的超参数设置.COPILOT 的超参数设置来源于经验结果.因此,通过进行更全面的实验调优可能会得到更佳的超参数设置,从而改善模型的性能.此外,本文所提方法在数据预处理阶段并未根据目标项目数据分布对源项目数据集进行筛选.针对 CPDP 任务

中存在的类别不平衡问题,Li 等人[35]提出一种基于数据选择和抽样的方法来解决上述局限性.因此,下一步研究工作将考虑设计有效的源项目选择策略来剔除那些与目标项目数据分布不一致的源项目,以进一步提升域适应模型的预测性能.

- 结构有效性: 结构有效性威胁首先来自使用的评估指标.在本文使用的评估指标之外,也可以使用其他评估指标(如 AUC)从不同角度验证模型的有效性.其次,本文所使用的数据集是从开源软件项目中挑选的八个 Java 项目.未来将采用更丰富的指标和更多的编程语言来提升 COPILOT 的可扩展性.

## 5 总结与展望

缺陷类别预测对提高软件开发人员进行缺陷定位及修复工作的效率是非常重要的.本文提出一种基于多源域适应的缺陷类别预测方法 COPILOT.该方法首先利用对抗训练来计算来自目标域的缺陷样本属于每个源域的概率.接着,COPILOT 选择加权最大均值差异 WMMD 作为注意力机制来进一步地最小化不同源域和目标域特征之间的表示距离,以减少域不相关性对模型性能的影响.本文选择 8 个来自 STONESOUP 计划的开源软件项目构建了缺陷数据集,并在该数据集上将 COPILOT 与多种来自软件工程和多源域适应领域的基线方法进行性能对比分析.结果验证了 COPILOT 可以减小不同源项目和目标项目数据集之间的数据分布差异,从而实现不同软件项目之间知识的有效迁移以提升跨项目场景下缺陷类别预测模型的分类性能.同时,本文所提出的模型代码已上传至开源仓库以方便后续研究.因此,在面向特定项目的实际软件开发维护场景中,软件实践者可以首先利用收集到的源项目数据集以及特定目标项目数据集对本文所提出的模型 COPILOT 进行进一步地精调,并将训练好的模型对特定项目中的缺陷实例进行类别预测.未来工作中将考虑在更多的编程语言以及更丰富的缺陷类别中扩展本文所提出模型.此外,本文针对函数级代码进行缺陷类别预测,因此在细粒度的代码表示(如语句级代码)中预测具体的缺陷类别也是重要的研究课题.

**附中文参考文献:**